\documentclass[manuscript]{aastex}
\usepackage[utf8]{inputenc}
\usepackage{graphicx}
\usepackage{amsmath} 
\usepackage{wasysym} 
\usepackage{bm}  

\newcommand{\up}[1]{\ifmmode^{\rm #1}\else$^{\rm #1}$\fi}

\newcommand{\arcd}{\ifmmode^{\circ}\else$^{\circ}$\fi}
\newcommand{\arcm}{\ifmmode{'}\else$'$\fi}
\newcommand{\arcs}{\ifmmode{''}\else$''$\fi}

\slugcomment{\today{} version}

\shorttitle{Double mode RR Lyr stars}
\shortauthors{Poleski}

\begin{document}


\title{Double-Mode Radial Pulsations among RR Lyrae Stars}

\author{Rados\l{}aw Poleski\altaffilmark{1}}
\email{poleski@astronomy.ohio-state.edu}

\altaffiltext{1}{Department of Astronomy, Ohio State University, 140 W. 18th Ave., Columbus, OH 43210, USA}


\begin{abstract}
Double-mode RR Lyr type stars are important for studies of properties of horizontal-branch stars. 
In particular, two periods coupled with spectral properties give a mass estimate that is independent of evolutionary models. 
Here, we present 
59 new Galactic double-mode RR Lyr stars found in the LINEAR survey data with the fundamental radial mode and the first overtone exited (RRd stars).
These stars may be useful for constraining the mass-metallicity relation for field horizontal-branch stars.
Also, new RRd stars found in the LMC by EROS-II are verified. 
We present the updated Petersen diagram and the distribution of the fundamental mode periods. 
Comments on selected variable stars from LINEAR and LMC EROS-II surveys are also presented, 
including very rare objects: the third known mode-switching RR Lyr and Cepheid pulsating simultaneously in three radial modes. 
\end{abstract}


\section{Introduction} 

The main parameter governing stellar evolution is mass; however, mass is hard to measure, 
as only a small portion of stars are members of eclipsing binaries for which mass may be measured directly.  
This especially affects horizontal-branch stars 
because their radii were much larger during previous phases of stellar evolution
{ and the chance of observing an eclipse is smaller}. 
Therefore, indirect methods must be used to estimate masses of horizontal branch stars.
One of such methods is based on properties of double-mode RR Lyr stars. 
RR Lyr type stars typically pulsate in fundamental mode (henceforth F; subtype RRab). 
The first-overtone pulsations (1O; subtype RRc) are less frequent. 
Here, we focus on subtype RRd stars that pulsate simultaneously in both modes (F/1O) with period ratio of $\approx 0.745$.
In some stars, the second overtone mode (2O; subtype RRe) is observed, but for monoperiodic stars, we are not able to distinguish between RRc and RRe type stars. 
We can statistically distinguish the two types, 
as RRe stars should have shorter periods and lower amplitudes \citep{alcock96}, but see discussion in \citet{clement00}.
Second-overtone pulsations are also found in double-mode objects \citep[e.g.][]{poretti10} with F/2O period ratios of $\approx0.59$. 
Around a dozen or so such objects are known. 
Properties of double-mode radial pulsators are investigated on the so-called Petersen diagram,  which presents period ratio versus longer period. 

Periods of double-mode stars coupled with spectroscopic properties give mass estimates \citep{popielski00,bragaglia01}. 
Until now, there have not been enough Galactic field stars with mass measured this way to allow statistical studies, most importantly deriving mass-metallicity relation for the horizontal-branch stars.
In order to facilitate this, 
we performed a search for double-mode RR Lyr stars in the publicly available dataset of the Lincoln Near-Earth Asteroid Research (LINEAR) survey data \citep{palaversa13}. 
Our search revealed 59 new field RRd stars,  more than doubling the number of objects of this type\footnote{
After this paper was submitted, \citet{drake14} claimed the discovery of 502 field RRd candidates. 
The reliability of this sample needs to be verified, as we do below with the EROS-II candidates, because they are claimed based on detection of a single period only. 
A significant number of the \citet{drake14} candidates have a dominant period beyond the range typically observed for RRd stars {\it i.e.,} from $0.34~{\rm d}$ to $0.43~{\rm d}$. 
Thus, we suspect false positives to be present in this sample. 
}. 
This search is described in next section. 
The reliability of the Large Magellanic Cloud (LMC) RRd stars found in the second phase of the Exp$\acute{\text{e}}$rience pour la Recherche d'Objets Sombres (EROS-II) project data is verified in Section~3.
We also comment on the claimed 
F/2O pulsators found in  the literature (Sec.~4.). 
The results are discussed in Section~5.
In the appendices, we present comments on other LINEAR and EROS-II variable stars of other types.

\section{Double mode RR Lyr stars in LINEAR data} 

The LINEAR survey operated between 1998 and 2009 using two $1~{\rm m}$ telescopes. 
Each of the mosaic cameras had a $2~{\rm deg^2}$ field of view with a $2.25\arcs$ pixel scale. 
Observations were collected without a filter and sometimes in nonphotometric conditions. 
The details of the instrumentation and the observing strategy can be found in \citet{sesar11}. 
On average 250 epochs were collected per field, and the number rises to 500 for targets close to  the Ecliptic plane. 
Such a dataset is suitable for a search for periodic variable stars. 
The catalog of such objects was presented by \citet{palaversa13}.
It includes more than 7000 stars, most of which are RR Lyr pulsators.
There are 2923 RRab and 990 RRc stars. 

We performed the search for other radial modes in 
RR Lyr type stars. 
The light curves of RR Lyr stars were $3\sigma$ clipped in order to remove obvious outliers (most probably caused by the nonphotometric conditions for some epochs). 
The periodograms were calculated using a discrete  Fourier transform.
Because of the strong daily aliases, up to the four highest peaks for each star were considered for further analysis if their amplitudes were larger than $95\%$ of the highest peak. 
The photometry was phased with each trial period; up to the fourth-order Fourier polynomial was subtracted from RRc star light curves, and up to the eighth-order for RRab stars.
Then, the periodograms were calculated for the prewhitened light curves, and once more up to the four highest peaks were considered.
The additional constrain was that the selected period could not be a daily alias of the previously found period. 
For each star, we checked which combination of periods gave smallest $\chi^2$ of the residuals. 
For 47 stars, the period ratio was similar to the ones found in other RRd stars. 
An additional 13 stars were selected among the stars for which a period ratio of $\approx 0.745$ resulted in $\chi^2$ only slightly higher than the smallest value. 
{ The smallest signal-to-noise ratio for accepted periods was 4.5.} 
Final estimates of periods and uncertainties were found using multiharmonic analysis of variance \citep{schwarzenbergczerny91,schwarzenbergczerny96} on disentangled light curves.

\begin{figure}
\epsscale{.79}
\plotone{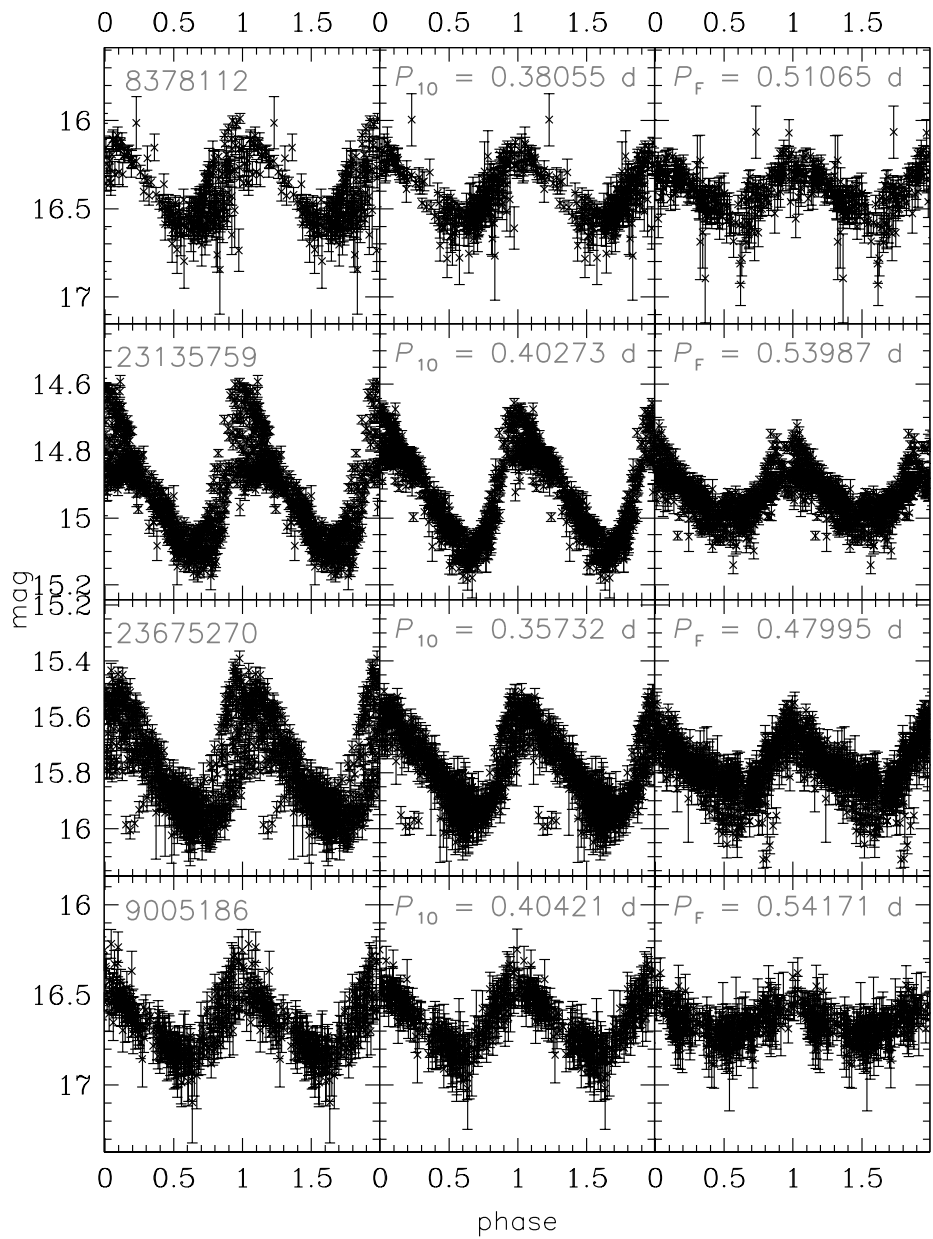} 
\caption{Randomly chosen light curves of RRd stars.
Each row represents one star. 
First column presents raw light curve phased with 1O period and gives LINEAR ID of the star.
Second and third column present disentangled light curves phased with 1O and F periods, respectively, and gives the periods of each star.
\label{fig:lc}}
\end{figure}

All of the stars that we identified, except one were classified by \citet{palaversa13} as RRc pulsators.
We performed an additional search in which stars marked as ''other'' by \citet{palaversa13} were examined.
This resulted in one candidate star that was examined in detail, and an RRd interpretation was confirmed. 
No double-mode pulsator with the 2O mode excited was found. 
The light curves of RRd stars were prewhitened with both modes, and the periodograms were calculated once more, but we did not find any additional radial mode. 
For four stars, our selection of periods is not definite, and it is possible we selected aliases of the true periods 
(LINEAR IDs:
2251043, 
4282396, 
9142645, and
19533186). 
{ 
In order to ensure our detections are correct, we have checked the variability of these stars in publicly available 
Catalina Real-time Transient Survey archive \citep{drake09}. 
The retrieved data were obtained mainly with the 0.7-m Catalina Sky Survey Schmidt Telescope, with some contributions from
1.5-m Mount Lemmon Survey Cassagrain Telescope. 
Both periods of the four stars were confirmed in Catalina photometry. 
}
We searched for previous detections of our RRd stars using SIMBAD database and found two counterparts:
LINEAR ID 23713423 = J10595715+3750203 \citep{sokolovsky09} and
LINEAR ID 13905365 = J15160921+3200073 \citep{bernhard12}.
We present the LINEAR ID, common name (if available), sky coordinates as well as both periods in Table~\ref{tab:res}.
The example light curves are presented in Figure~\ref{fig:lc}.

\section{Double mode RR Lyr stars in the EROS-II LMC data} 
Recently, the EROS-II project presented a catalog of 117 thousand variable stars \citep{kim14}. 
They were selected using a machine-learning algorithm based on photometric data collected 
from the LMC between 1996 and 2003 using 1 m telescope. 
Exposures were simultaneously taken in two nonstandard bands thanks to a dichroic beamsplitter.
There are 55 thousand newly discovered variable stars in this catalog.
The training set for the classifier was  mainly based on the catalog of variable stars provided by the third phase of the Optical Gravitational Lensing Experiment (OGLE-III). 
In case of RR Lyr stars, \citet{soszynski09rrlyrlmc}  data were used.
The \citet{kim14} catalog contains 988 stars claimed to be RRd type stars. 
Out of these, 495 stars were previously published \citep{alcock97,soszynski09rrlyrlmc}. 
The remaining 493 stars may significantly increase the sample of about a thousand 
RRd stars known in the LMC. 
However, further investigation of these stars poses severe challenges; 
\citet{kim14} presented only a single period for each star, and thus these objects cannot be placed on the Petersen diagram and analyzed.
In this section, we 
verify the existence of the second pulsation mode in these stars.  

First, we checked the available OGLE photometry. 
The fourth phase of the OGLE project presented a catalog of variable stars 
towards the South Ecliptic Pole 
\citep{soszynski12}, which falls within the tidal radius of the LMC, 
but was not 
cross-matched with the EROS-II catalog by \citet{kim14}. 
We found 21 common objects, with 18 classified as RRd stars by \citet{soszynski12}. 
The sample of RRd stars from \citet{kim14} contains 68 objects that are in the OGLE-III catalog of variable stars and are classified differently. 
There are 33 RRc stars, 30 RRab stars, three classical Cepheids, and two contact binaries. 
We obtained photometry of these stars from 
\citet{soszynski09rrlyrlmc}, \citet{soszynski08}, and \citet{graczyk11}. 
Both the third and the fourth phase of the OGLE project observed with the 1.3 m Warsaw telescope at the Las Campanas Observatory
(Chile) 
with large CCD cameras. 
Almost all the data were collected with an $I$-band filter, and only these are analyzed here. 
For more details, see \citet{udalski08red} and \citet{soszynski12}. 
The number of epochs for an individual star varies between 115 and 1016 with median of 370. 
The median detection limit for the secondary period was $0.039~{\rm mag}$ (at signal-to-noise ratio of 4.5). 
Each star was examined in detail, and in only one case 
(OGLE-LMC-RRLYR-22133 = lm0221n9969) was a bona fide RRd star revealed. 

\begin{figure}[!htb]
\includegraphics[bb=281 16 580 743,clip,angle = 270,width=.99\textwidth]{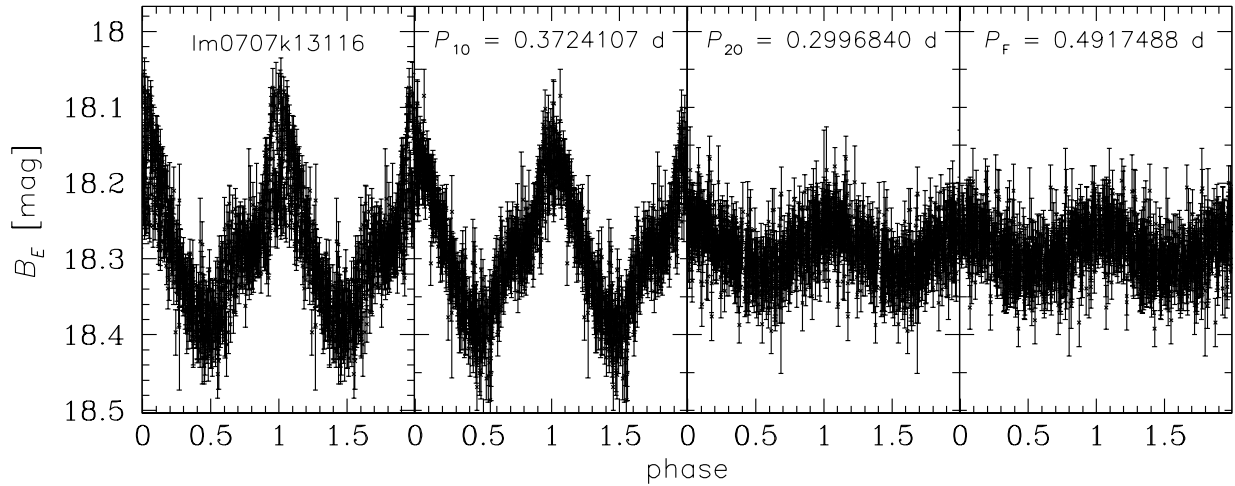} 
\caption{ Light curve of newly found triple-mode Cepheid lm0707k13116.
First panel shows raw data phased with most significant period. 
Each of the three other panels presents data phased with period indicated after removing two other signals. 
\label{fig:cepF1O2O}} 
\end{figure}

Second, the EROS-II nonstandard $B$-band light curves from \citet{kim14} were examined. 
There are between 163 and 630 epochs (median of 480). 
The median detection limit for the secondary period was higher than in the case of the OGLE data, and was equal to $0.046~{\rm mag}$. 
We were looking for particular type of variability, and the number of objects investigated was low enough to allow a case-by-case study. 
Our investigation confirmed 350 RRd stars, bringing the total number of these objects in the LMC to 1354. 
The full list of confirmed objects is presented in the online material, while a partial example is shown in Table~\ref{tab:eros_newRRd}. 

Except the bona fide RRd stars, we found secondary periods in eight other objects (Table~\ref{tab:eros_other}).
The most important one is star lm0707k13116 ($\alpha = 05^{\rm h}25^{\rm m}47.87^{\rm s}$, $\delta = -73^{\circ}02'15.0''$), which was identified by \citet{marquette09} as a 1O/2O classical Cephied (ID: J052547-730214). 
We revealed that it has not only the first two overtones excited, but also the fundamental mode is clearly present (Figure~\ref{fig:cepF1O2O}).
This star is very similar to another F/1O/2O Cepheid, OGLE-LMC-CEP-1378 \citep[$P_{\rm F}\approx0.515~{\rm d}$, $P_{\rm 1O}\approx0.385~{\rm d}$, and $P_{\rm F}\approx0.309~{\rm d}$;][]{soszynski08}.
Note that only 10 triple-mode Cepheids were previously identified \citep{moskalik04,soszynski08,soszynski10cepsmc,soszynski11cepblg}. 
The presence of the three radial modes in one star gives strong constrains on pulsation model \citep{moskalik05}. 
The interpretation of seven other stars identified in Table~\ref{tab:eros_other} is less clear. 
Period ratios for some of them are close to expected for RRd stars, but their position on the Petersen diagram differs from expected. 
They may be nonradial pulsators \citep{dziembowski99}. 
Some of the seven stars may be either $\delta$ Sct type pulsators or classical Cepheids. 
We note that there is no star that has both a 1O and 2O mode present and is securely identified as an RR Lyr type star \citep{olech09}. 
The observational distinction between the shortest-period Cepheids and the longest-period RR Lyr and $\delta$ Sct stars should be addressed. 

\begin{figure}
\begin{center} 
\includegraphics[bb=40 308 570 738,width=.8\textwidth]{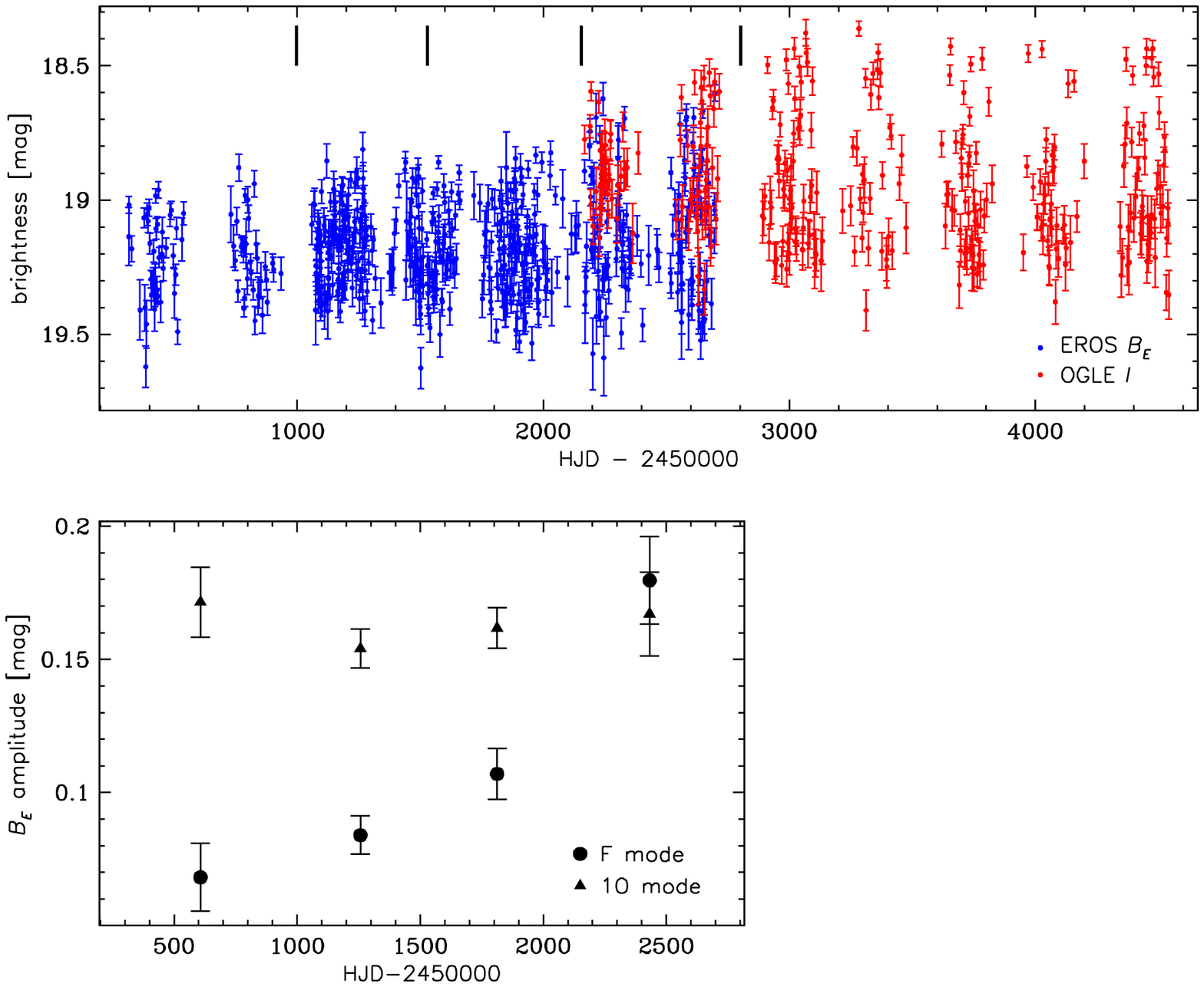} 
\end{center}
\caption{Light curve evolution of mode switching RR Lyr lm0575m5221 = OGLE-LMC-RRLYR-13308. The upper panel presents EROS-II $B_E$- and OGLE $I$-band photometry. Black marks show epochs that were used to divide data into separate chunks. The lower panel presents amplitude changes of both modes. The time axis of both panels are aligned. \label{fig:mode_switch}} 
\end{figure}

\begin{figure}
\begin{center}
\includegraphics[bb=47 48 472 742,angle=270,width=.8\textwidth]{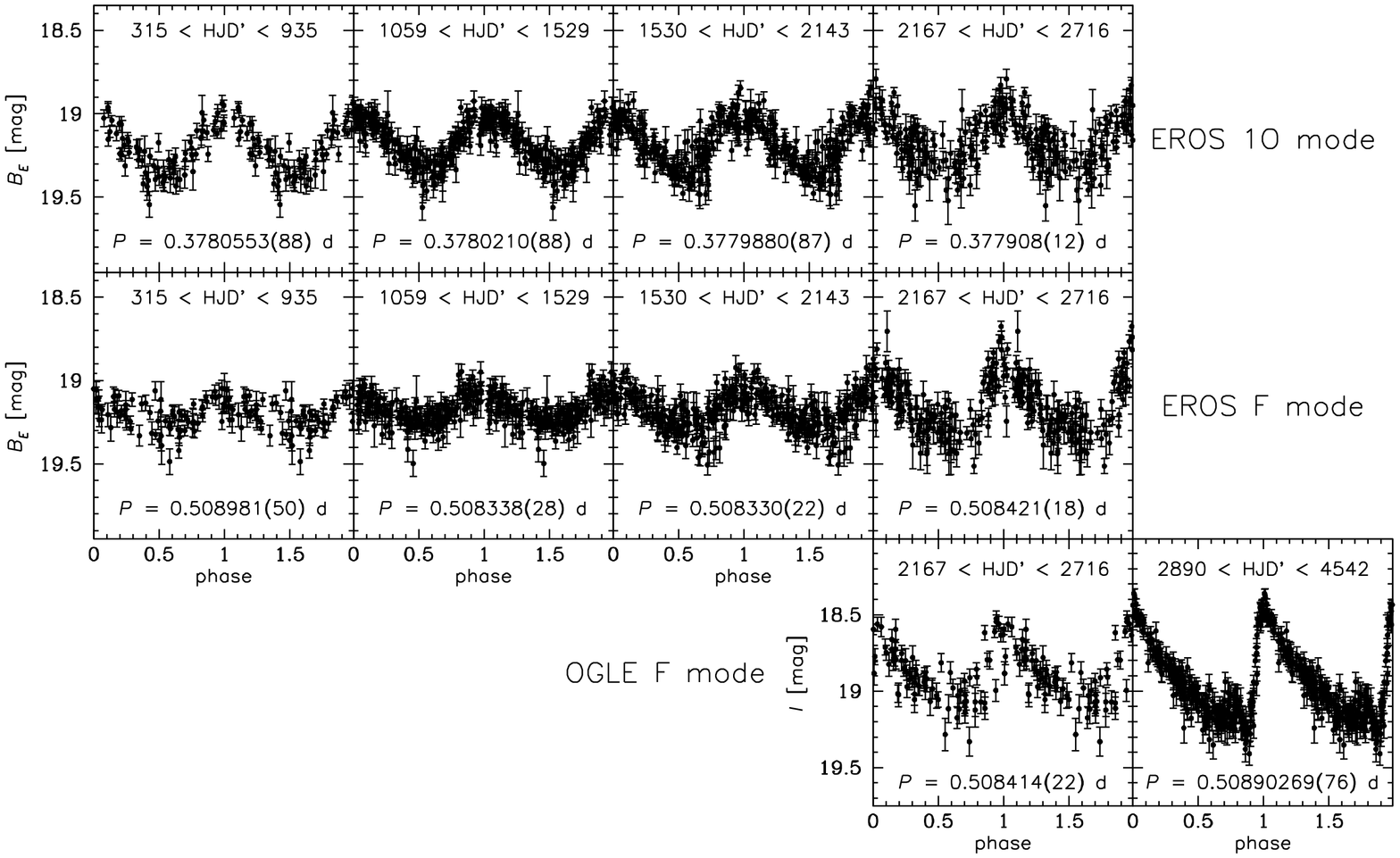}
\end{center}
\caption{Evolution of phased light-curves of mode switching RR Lyr lm0575m5221 = OGLE-LMC-RRLYR-13308.
Disentangled EROS data are shown for 1O mode (top row) and F mode (middle row) separately.
Bottom row shows F mode in OGLE data.
Each column presents separate chunk of data (see Figure 3) which time interval and derived period are indicated in each panel. \label{fig:mode_switch2}}
\end{figure}

We checked the OGLE-III classification of RRd stars confirmed in the EROS-II data. 
Two discrepant objects were found. 
The first has already been mentioned: 
OGLE-LMC-RRLYR-22133 = lm0221n9969, which turned out to be an RRd not recognized in the OGLE-III catalog.
The second object is more interesting: lm0575m5221 = OGLE-LMC-RRLYR-13308 ($\alpha = 05^{\rm h}22^{\rm m}43.55^{\rm s}$, $\delta = -71^{\circ}10'48.2''$). 
The light curve is presented in Figure~\ref{fig:mode_switch}. 
We see that in EROS-II data (points with uncertainties larger than $0.15~{\rm mag}$ were removed) the amplitude of variability increases with time. 
The OGLE data show a small increase of amplitude at the beginning, followed by period of steady amplitude. 
EROS-II data were divided into four separate chunks and checked for periods present. 
In each of them, we independently found two periods about $0.3780~{\rm d}$ and $0.5085~{\rm d}$. 
Their ratio is $0.7435$; thus, the first is 1O mode and the second is F mode. 
The amplitudes of both modes as a function of time are presented in lower panel of Figure~\ref{fig:mode_switch}. 
We see a constant amplitude of 1O mode and a fast rising amplitude of the F mode. 
The end of the EROS-II data overlaps with the first two years of the OGLE-III data. 
However, in these data we cannot find the 1O mode, even though some changes in the light curve are seen. 
The amplitudes of RR Lyr stars are smaller in longer wavelengths, which probably makes the 1O amplitude below the detection limit in the $I$-band.  
\citet{soszynski09rrlyrlmc} classified this object as RRab, but we now see this is in fact mode-switching star. 

Based on our analysis, we can draw conclusions on the reliability of the selection of RRd stars in \citet{kim14}. 
Out of 988 stars listed 863, ({\it i.e.,} $87\%$) are confirmed as RRd either here or in previous works. 
There are 1354 RRd stars known in the EROS-II LMC footprint, and $64\%$ of them were revealed. 
However, in the sky area covered by both the EROS-II and the OGLE-III surveys, \citet{kim14} found only $50\%$ of the RRd pulsators. 
For each star, they also provided the probability that variability class is assigned properly (class probability),  
which was estimated based on statistics returned by a random forest classifier run on EROS-II data \cite[for details see sec. 3.4.1 in][]{kim14}. 
We found that mean class probability for confirmed objects is $0.86$, while for unconfirmed objects it is $0.47$. 
This means that, in general, objects with higher class probability are more likely RRd stars, 
but for a significant number of false positives the class probability is a high number.

\section{ F/2O RR Lyr stars in the OGLE-III LMC data} 

\citet{chen13} presented the analysis of the catalog of the RR Lyr stars in the LMC. 
The input data were taken from the OGLE-III catalog \citep{soszynski09rrlyrlmc}. 
One of the results presented by \citet{chen13} was the identification of additional frequencies in eight stars, four of which were claimed to be F/2O pulsators. 
Our analysis of the claimed periods reveals significant problems with the interpretation presented by \citet{chen13}. 

There are two stars with secondary periods close to $1~{\rm d}$. 
\citet{chen13} denoted them as reliable detections based on the folded light curve.
There are two more stars with secondary periods close to $0.5~{\rm d}$. 
The unphased light curves of these stars present significant long-term trends in the data. 
These stars have OGLE identifications: 
OGLE-LMC-RRLYR-07071, 
OGLE-LMC-RRLYR-16758, 
OGLE-LMC-RRLYR-13185, and 
OGLE-LMC-RRLYR-17333\footnote{ 
\citet{chen13} have not made public full Table~1., which was suppose to give conversion between their IDs and the OGLE ones.
We found the OGLE identification number based on the description of stars selection procedure, which was further verified by exact matching of periods.}
.
{ We prewhinted each light curve with the most significant period. 
The secondary periods claimed by \citet{chen13} were found in each case. 
From the prewhinted light curve of the first star, we removed the moving average in $100~{\rm d}$ wide bins. 
The recalculated periodogram did not show any sign of a secondary period. 
The second star gets fainter each year by $0.059~{\rm mag}$. 
\citet{soszynski09rrlyrlmc} presented the photometry from the second and the third phase of the OGLE project. 
The alignment of the two data sets was imperfect for this star, thus we used only the OGLE-III data. 
Removing the $0.059~{\rm mag/yr}$ slope from these data also eliminates the secondary period. 
The third star shows changes in mean brightness only in the OGLE-II data, and only for them the secondary period is present. 
We removed the moving average in $200~{\rm d}$ wide bins and did not see the secondary period afterwards. 
The same procedure was applied to the whole light curve of the fourth star and the secondary period also vanished. 
}

One more star claimed by \citet{chen13} to be a F/2O pulsator (OGLE-LMC-RRLYR-08575) 
has two periods, which are daily aliases of one another.
As such, it cannot be considered a F/2O pulsator without independent observational confirmation. 
\citet{chen13} also claimed a detection of secondary period for OGLE-LMC-RRLYR-13058, which resulted in period ratio of 0.693. 
Following \citet{soszynski09rrlyrlmc} we note that this star is a blend of an RR Lyr variable and an eclipsing binary. 
We checked that after subtracting the eclipsing variability, the putative secondary pulsation variability disappears. 

Out of the eight RR Lyr stars identified by \citet{chen13} as showing two periods, we are left with two objects, which are 
OGLE-LMC-RRLYR-12967 (longer period $P_L=0.5769~{\rm d}$, period ratio $P_S/P_L = 0.5762$) and 
OGLE-LMC-RRLYR-13231 ($P_L=0.5992~{\rm d}$, $P_S/P_L = 0.7758$). 
The former star has a secondary period close to $1/3~{\rm d}$ but it does not correspond to the maximum of window function and thus should be real and the star is strong candidate for the F/2O pulsator.
The secondary period for the latter object was identified already by \citet{soszynski09rrlyrlmc}.
Its period ratio is significantly larger than any other RRd star \citep[$0.726-0.750$][]{soszynski11rrlyrblg}.

\section{Discussion} 

We presented a list of 59 new RRd stars in the Galaxy field. 
Previously known sample contained 57 field objects \citep[][and references therein]{cseresnjes01,sokolovsky09,soszynski09rrlyrlmc,soszynski10rrlyrsmc,wils10,bernhard12,chadid12,mcculsky08}, six of which are in front of the Magellanic Clouds (globular clusters and dwarf spheroidal galaxies RRd stars were excluded). 
There are also 80 such stars in the OGLE-III Galactic bulge fields \citep{soszynski11rrlyrblg}. 
This shows that our sample significantly increases the sample of Galactic RRd stars. 
The visual magnitudes of our newly found RRd stars are between 14 and 17, which
makes them relatively good targets for follow-up spectroscopy. 
Such observations will allow deriving masses of statistically significant number of field horizontal branch stars \citep{bragaglia01}.

\begin{figure}
\plotone{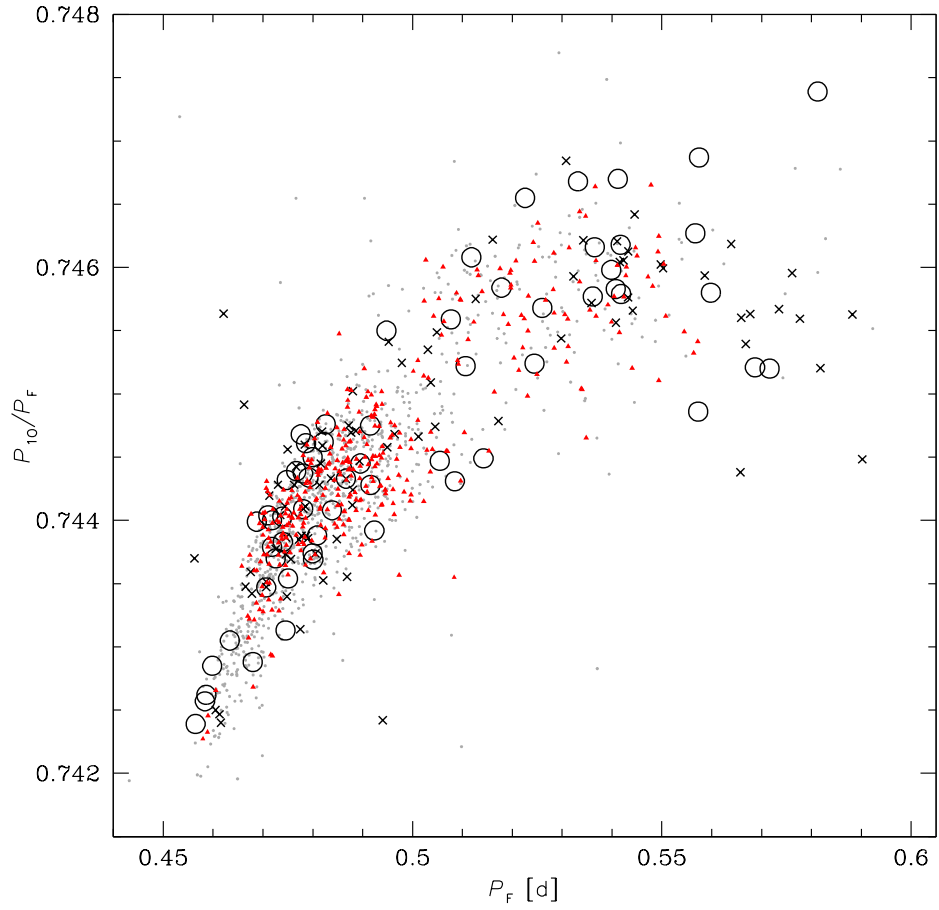}
\caption{Petersen diagram for RRd stars. 
Large circles mark stars we found in the LINEAR photometry.
The other field Galactic stars are marked by smaller black crosses (36 bulge RRd stars fall outside the plotted region).
Previously known LMC objects are shown by gray dots for comparison.
New LMC RRd stars are marked by red triangles. 
\label{fig:Peter}} 
\end{figure}

The Petersen diagram 
of the newly found RRd stars as well as that of the previously known ones is presented in Figure~\ref{fig:Peter}. 
One can see that newly found objects fall in the region, in which previously found variables of this type are situated. 
Based on models presented by \citep{soszynski11rrlyrblg}, we estimate that the metallicities $[{\rm Fe/H}]$ are between $-1.7$ and $-1.1$, except for the star with the highest period ratio (i.e., LINEAR ID 22316675, $P_{\rm F} = 0.5813~{\rm d}$, $P_{\rm 1O}/P_{\rm F} = 0.7473$). 
This star should have metallicity of $[{\rm Fe/H}]\approx-2$.

\begin{figure}
\plotone{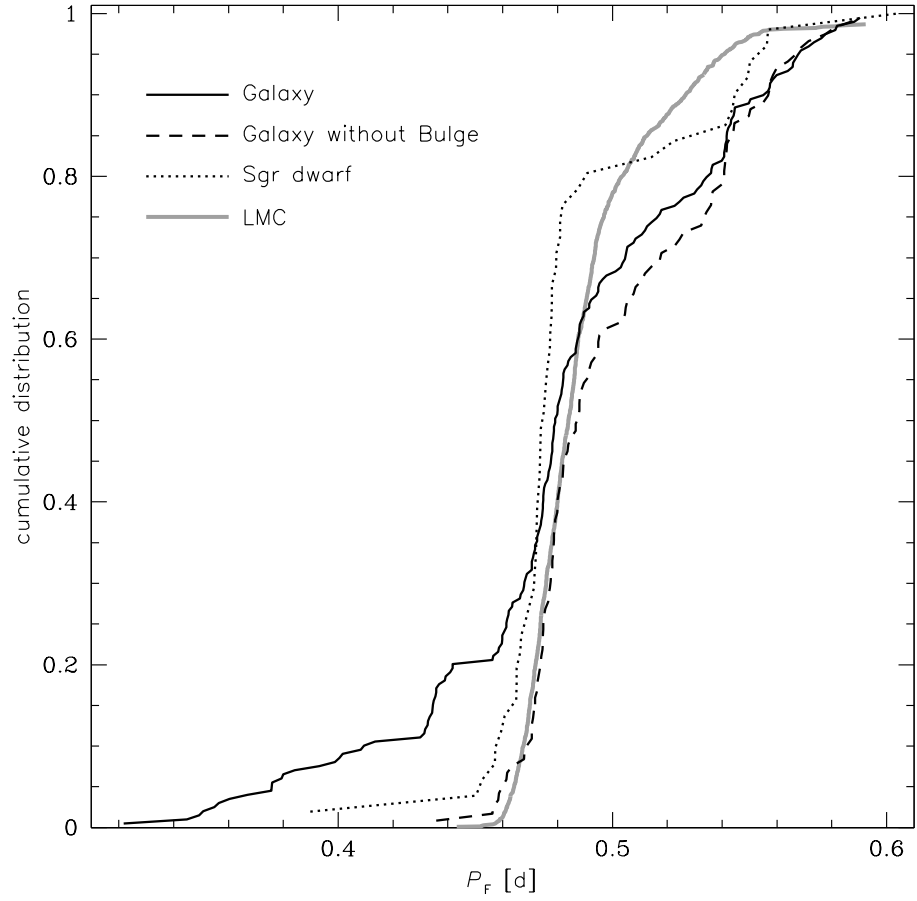}
\caption{Cumulative distribution of fundamental mode periods of RRd stars. 
\label{fig:perdist}}
\end{figure}

The period distribution of RRd stars depends on many factors like star formation history, metallicity of the environment, and possibly on resonances of F, 1O, and 2O modes \citep[$2\omega_{\rm 1O} = \omega_{\rm F} + \omega_{\rm 2O}$;][]{kovacs88}. 
We present the  cumulative distribution of the F periods in Figure~\ref{fig:perdist}. 
The distribution of the Galactic RRd stars is given by the solid line, and the dashed line illustrates the same distribution without taking the Bulge stars into account. 
\citet{wils10} analyzed this distribution using 89 Galactic stars. 
They noted the period gap between $0.52~{\rm d}$ and $0.53~{\rm d}$. 
Using a sample more than twice larger, we cannot confirm the existence of this gap. 
The highest number of Galactic RRd stars have F periods between $0.46~{\rm d}$ and $0.49~{\rm d}$. 
The number sharply rises around periods of $0.44~{\rm d}$ (only Bulge fields) and $0.54~{\rm d}$. 
We also presented the same distribution for 51 stars located in the Sgr dwarf spheroidal galaxy \citep{cseresnjes01,soszynski11rrlyrblg}. 
It clearly shows binomial distribution \citep{wils10}, with most of the stars in the period range $0.46-0.48~{\rm d}$ and a smaller number concentrated between $0.54~{\rm d}$ and $0.56~{\rm d}$. 
None of the above claims should be significantly affected by detection biases connected to the daily aliases in the photometry. 

For only three newly found RRd stars F mode amplitude is larger than 1O mode amplitude.
One of these stars was found in LINEAR data (ID 23713423 $P_{\rm F} = 0.475~{\rm d}$) and two in EROS-II data
(lm0355k17105 $P_{\rm F} =  0.458~{\rm d}$, lm0417m11567 $P_{\rm F} =  0.466~{\rm d}$).
These stars are at the short tail of period distribution.
\citet{oaster06} also found that stars with the F mode amplitude larger than the 1O mode amplitude are grouped at the shorter periods using larger sample of such stars.

We have verified in detail the RRd stars presented by \citet{kim14}.  
These stars are marked in the Petersen diagram (Figure~\ref{fig:Peter}). 
There are no significant differences between the locations of previously known LMC RRd (gray dots) and newly revealed stars (red dots). 
The distribution of fundamental periods presented in Figure~\ref{fig:perdist} does not show any discontinuities. 
We confirmed most of the analyzed stars, but many false positives were also identified. 
We suggest that stars should not be classified as RRd, {\it i.e.,} double radial mode pulsators, based on a single period only. 
Our investigation of the EROS-II variables reveled two very rare examples of variable stars: 
one triple-mode classical Cepheid and one RR Lyr that switched from simultaneous F and 1O to pure F mode pulsations. 

The RR Lyr star switching pulsation modes is only the third such star known \citep{kaluzny98,soszynski14a} 
and the one for which the largest observing material during mode changing was collected\footnote{
After this paper was submitted, \citet{drake14} claimed the discovery of six new mode-switching stars. 
\citet{drake14} presented light curve of one of these stars and this is certainly not a star switching between two radial modes as suggested by these authors,"  
One example presented in the paper is certainly not a star switching between two radial modes as suggested by \citet{drake14}, 
because the periods are very similar instead of being close to ratio of $0.745$, $0.80$ or the product of the two. 
The cause of significant amplitude changes and small period changes in this particular object is most probably a Blazhko-like effect. 
For similar example, see OGLE-BLG-RRLYR-07011 in \citet{soszynski11rrlyrblg}. 
We do not include six \citet{drake14} objects into mode switching RR Lyr stars sample. 
}. 
Our analysis of the EROS-II data showed that 1O period was changing at a rate of $-2.88(50)\times10^{-5}~{\rm d/yr}$. 
In contrast, the F period  was first decreasing from 
$0.508981(49)~{\rm d}$ to 
$0.508375(12)~{\rm d}$ during $\approx650~{\rm d}$, and 
it was constant later on. 
OGLE data indicate an even higher value of 
$0.5089004(7)~{\rm d}$. 
In the case of OGLE-BLG-RRLYR-12245 \citet{soszynski11rrlyrblg} found a period change rate of fundamental mode of $6\times10^{-5}~{\rm d/yr}$. 
They neglected evolutionary changes as a cause of period change rate, which is in line with our findings. 
More detailed analysis of light curve of lm0575m5221 = OGLE-LMC-RRLYR-13308 should be possible with archival photometry of EROS-II, OGLE, and possibly MACHO projects. 
In the case of the EROS-II data, we suspect the quality of the photometry can be improved. 

Based on a large sample of RRd stars found in the EROS-II data and the OGLE-III photometry collected in sucessive years, we can estimate the rate of mode-switching events. 
There are 497 stars common for both samples. 
Our search was sensitive only for stars which were classified as RRd by \citet{kim14} and differently by \citet{soszynski09rrlyrlmc}. 
Mean epochs of both surveys are separated by about $5.5~{\rm yr}$, thus the rate of mode-switching events seems to be an order of one in $2700{\rm yr}$.  
Microlesning surveys OGLE, MOA, MACHO and EROS-II have found and monitored $\approx1800$ RRd stars in the area of the Magellanic Clouds and the Galactic bulge. 
These stars have $\approx15~{\rm yr}$ long light curves collected with just a few different instruments setups, which makes this sample very well situated for a detailed search of mode-switching objects. 
Such a search should reveal about 10 objects of this type. 
Another $\approx45,000$ known RRab and RRc stars in this sky area can also be searched for mode-switching phenomena. 

During our analysis, we had to cope with daily aliases, and the ambiguity caused by them was not fully resolved. 
Similar problems were faced when we reanalyzed the periods detected by \citet{chen13}.
We note that special care has to be taken to check if the secondary periods of the RR Lyr stars are not caused by daily aliases.

\acknowledgments

I thank the anonymous referee, Marc Pinsonneault, Wojciech Dziembowski, and Andrew Drake for comments on the manuscript. 
The LINEAR, CSS, and EROS-II surveys are acknowledge for making their data available.
This research has made use of the SIMBAD database, operated at CDS, Strasbourg, France. 
CRTS and CSDR2 are supported by the U.S.~National Science Foundation under grant AST-1313422. The CSS survey is funded by the National Aeronautics and Space Administration under Grant No. NNG05GF22G issued through the
Science Mission Directorate Near-Earth Objects Observations Program. 

\appendix
\section{Comments on other LINEAR variables} 

The catalog presented by \citet{palaversa13} contains more than 7000 variable stars of different kinds. 
With growing observing capabilities, more and more variable stars are discovered, and their classification gets more and more time-consuming. 
In order to reduce the data from future time domain surveys such as GAIA or LSST, one has to find the automated algorithms that classify variable stars. 
Many of such algorithms require the training set of classified variables. 
Such a set should be as clean as possible. 
We have visually examined { selected} variables presented by \citet{palaversa13} and found that, for a significant number of them, the classification or periods can be corrected. 
We present our findings in Table~\ref{tab:other} with the primary goal of allowing training the algorithms on clearer samples. 
The first six columns of this table (LINEAR ID, coordinates, brightness, period, type) come from \citet{palaversa13}. 
The last two columns present period derived by us and the type of variability, which resulted from visual inspection of the light curve. 
For double-mode SX Phe or $\delta$ Sct pulsators, we also present their periods.
We note that seven dwarf novae with GCVS identifications were presented by \citet{palaversa13}.
In order to make access to their photometry easier, 
we list their cross-identifications:
1913172  = QZ Vir, 
6660642  = GZ Cnc,
10136638 = CR Boo,
17146068 = X Leo,
17301605 = RZ LMi,
17388042 = ER UMa, and
22107778 = KS UMa. 
Periods of $-9.9~{\rm d}$ correspond to objects for which reliable period could not be found. 
Below, we comment on the most common problems with \citet{palaversa13} classification and periods.

We found that for the most of SX Phe or $\delta$ Sct pulsators the light curve phased with provided period show the scatter larger than expected. 
The periods were recalculated, and it turned out that not enough significant digits of the period were presented. 
We found that \citet{palaversa13} truncated the periods to $10^{-6}~{\rm d}$ instead of rounding them and presenting more digits. 
For 82 stars classified as ''other'', with no period provided, we were able to found the period, which in our opinion well fits to the data. 

One more problem we found in the \citet{palaversa13} catalog are stars lying close to each other on the sky. 
As an example, we point our that there are 17 stars for which LINEAR IDs start with 1992, and all have periods in a narrow range of $0.246369~{\rm d} - 0.246372~{\rm d}$. 
This shows that their variability is caused by some bright star. 
Finding the real source of variability is very hard without the access to the raw images, thus we do not fully resolve this issue.
We point that similar problems are found with other stars. 

\section{Periods of eclipsing binaries in EROS-II catalog} 

\begin{figure}
\plotone{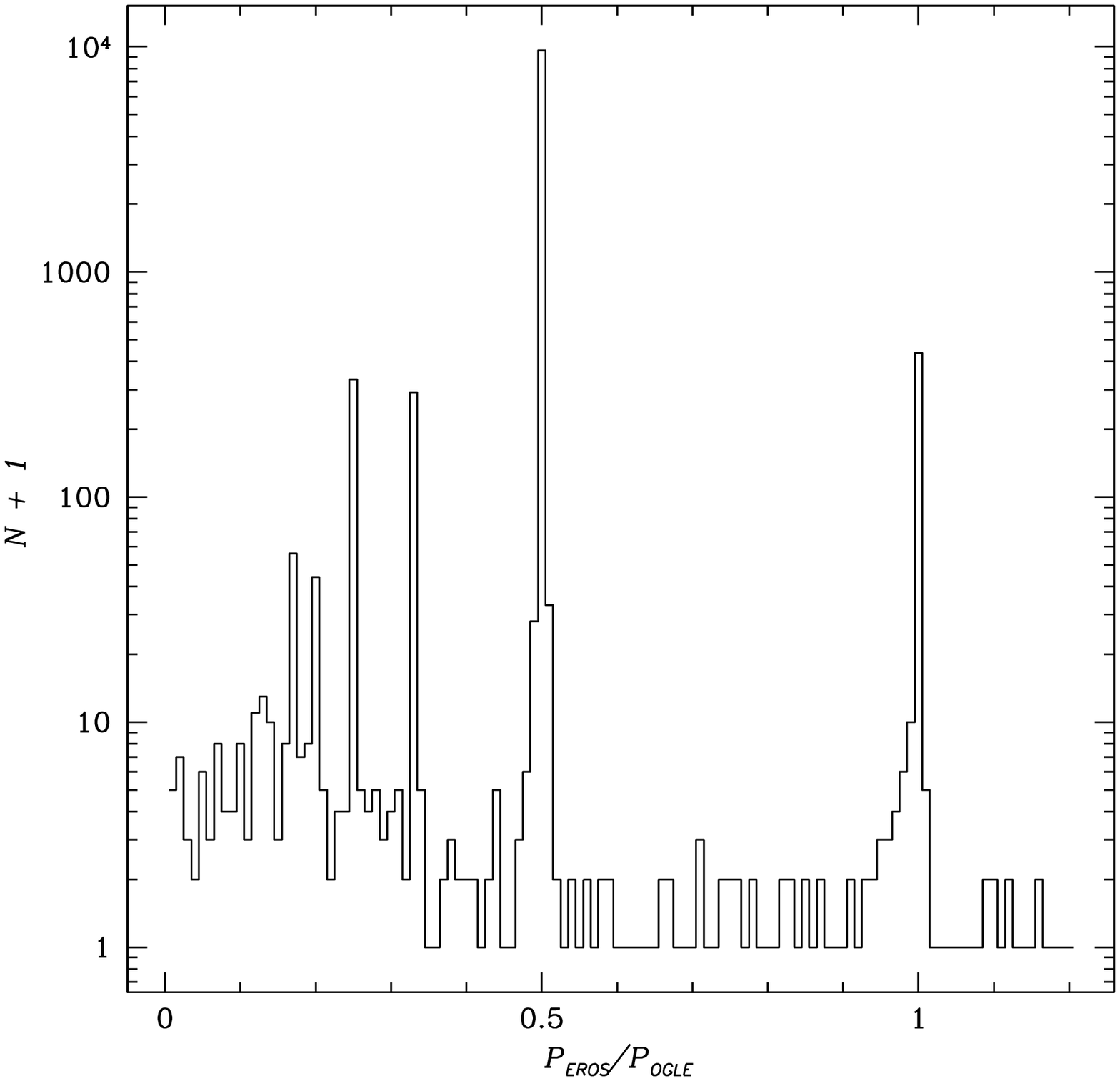}
\caption{Histogram of the period ratios in the EROS-II and the OGLE-III catalogs of eclipsing binaries. 
The bin size is $0.01$. 
There are 57 objects that fall beyond the $x$-axis range.
\label{fig:eroseb}}
\end{figure}

\begin{figure}
\plotone{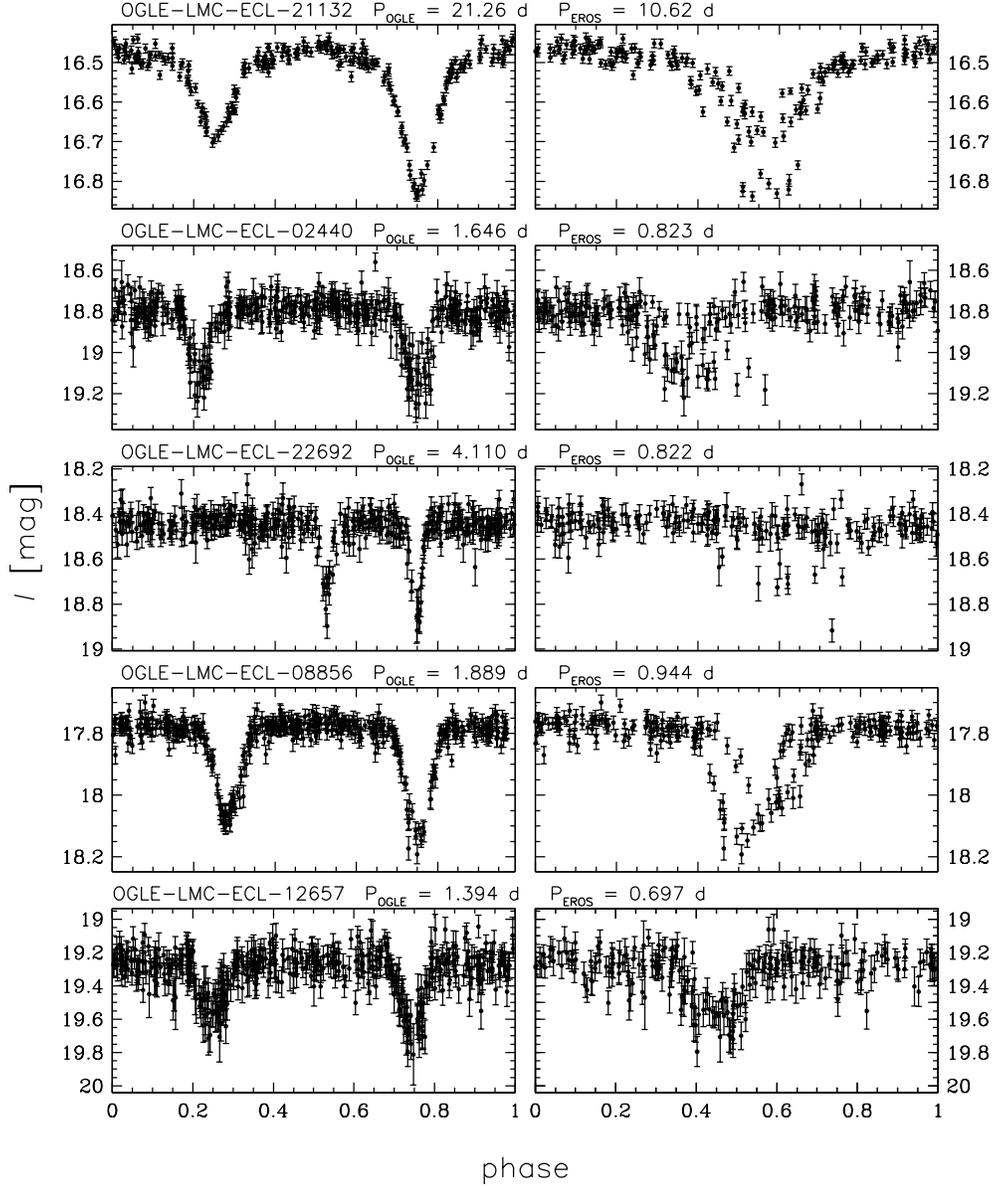} 
\caption{ Comparison of OGLE photometry for eclipsing binaries phased with periods provided by OGLE (left panels) and EROS-II (right panels) collaborations. 
Each row presents a separate star with OGLE ID and both periods given above the panels. 
The stars in last two rows were presented in Fig.~7 of \cite{kim14}. 
Reference epoch ($t_0$) was taken from \citet{graczyk11}, shifted by 1/4 of the period and used in both panels of each row. 
\label{fig:EBoe}}
\end{figure}

We have compared the periods for 
11,080 objects classified by both \citet[][ OGLE-III data]{graczyk11} and \citet{kim14} as eclipsing binaries.
This was done before EROS-II time-series photometry was made public,  
but we confirmed the results presented below using EROS-II data afterwards. 
The histogram of period ratios is presented in Figure~\ref{fig:eroseb}. 
Only  
$\approx 4\%$ 
of eclipsing binaries have period ratio close to unity.
For 100 randomly selected targets, we verified the claimed periods, and in each case the period provided by \citet{graczyk11} was correct. 
{ Example plots are shown in Figure~\ref{fig:EBoe}.}
The histogram of period ratios shows  not only the maximum expected at unity, but also other local maxima close to values of $1/2$, $1/3$, $1/4$, $1/5$, and $1/6$. 
The latter ones are  predominantly caused by binaries with eccentric orbits. 
As much as $87\%$ of eclipsing binaries have a period ratio in the range from $0.49$ to $0.51$. 
We attribute the large number of incorrect periods to the Lomb-Scarlage period search method used by \citet{kim14} that
fails for eclipsing binaries \citep[see, {\it e.g.,}][]{derekas07}.  
The light curves of eclipsing binaries have a nonsinusoidal shape--thus, fitting sinusoid model, as the Lomb-Scarlage periodogram does, gives periods that are not orbital periods. 
This problem is severe and affects 53,953 eclipsing binaries identified in \citet{kim14}, which is $36\%$ of their catalog of variable stars. 

Class probabilities for eclipsing binaries group around $0.4$ instead of unity. 
We found the cause of this discrepancy. 
\citet{kim14} used orbital periods to train the classifier, but the periods used as input for classifier were not orbital ones, as was shown above. 
For star with similar periods provided by \citet{graczyk11} and \citet{kim14}, we found median class probability of $0.77$. 
We expect that well-calculated periods would increase the efficiency of the classifier in classifying eclipsing binaries.

\begin{deluxetable}{lrrrrr}
\tabletypesize{\footnotesize}
\tablecaption{RRd pulsators found in LINEAR data\label{tab:res}}
\tablewidth{0pt}
\tablehead{
\colhead{LINEAR ID} & \colhead{Name} & \colhead{R.A.} &
\colhead{Dec.} & \colhead{$P_{\rm 1O}~[{\rm d}]$} & \colhead{$P_{\rm F}~[{\rm d}]$}
}
\startdata
658512   &        & $08^{\rm h}26^{\rm m}22.98^{\rm s}$ & $+28^{\circ}24'04.6''$ & $ 0.3526036 (10)$ & $ 0.4739116 (35)$ \\
737951   &        & $08^{\rm h}29^{\rm m}00.26^{\rm s}$ & $+37^{\circ}10'13.6''$ & $ 0.3570234 (11)$ & $ 0.4800741 (33)$ \\
4670587  &        & $08^{\rm h}36^{\rm m}11.13^{\rm s}$ & $+25^{\circ}34'11.9''$ & $ 0.3576988 (07)$ & $ 0.4808514 (18)$ \\
5514063  &        & $08^{\rm h}51^{\rm m}47.16^{\rm s}$ & $+07^{\circ}23'53.7''$ & $ 0.4003548 (15)$ & $ 0.5365345 (50)$ \\
5974119  &        & $09^{\rm h}04^{\rm m}04.50^{\rm s}$ & $+15^{\circ}26'42.0''$ & $ 0.3660638 (06)$ & $ 0.4915140 (40)$ \\
6182790  &        & $09^{\rm h}10^{\rm m}47.76^{\rm s}$ & $+50^{\circ}38'50.5''$ & $ 0.3405916 (17)$ & $ 0.4586306 (23)$ \\
6838366  &        & $09^{\rm h}20^{\rm m}39.43^{\rm s}$ & $+04^{\circ}54'36.6''$ & $ 0.3828368 (11)$ & $ 0.5142220 (32)$ \\
6958952  &        & $09^{\rm h}23^{\rm m}25.44^{\rm s}$ & $+05^{\circ}04'07.8''$ & $ 0.4747843 (28)$ & $ 0.3533945 (09)$ \\
6711303  &        & $09^{\rm h}27^{\rm m}30.68^{\rm s}$ & $+11^{\circ}07'38.2''$ & $ 0.4163646 (09)$ & $ 0.5574830 (38)$ \\
7337289  &        & $09^{\rm h}28^{\rm m}00.19^{\rm s}$ & $+06^{\circ}37'59.4''$ & $ 0.4041146 (08)$ & $ 0.5412042 (39)$ \\
7260181  & WY LMi & $09^{\rm h}30^{\rm m}23.26^{\rm s}$ & $+33^{\circ}53'10.6''$ & $ 0.3662646 (11)$ & $ 0.4923382 (26)$ \\
7279621  &        & $09^{\rm h}32^{\rm m}53.55^{\rm s}$ & $+37^{\circ}16'54.4''$ & $ 0.4154632 (13)$ & $ 0.5567090 (47)$ \\
7376901  &        & $09^{\rm h}34^{\rm m}11.67^{\rm s}$ & $+63^{\circ}19'59.3''$ & $ 0.3556595 (10)$ & $ 0.4775950 (32)$ \\
7519479  &        & $09^{\rm h}38^{\rm m}20.38^{\rm s}$ & $+20^{\circ}33'59.9''$ & $ 0.3510119 (08)$ & $ 0.4717880 (46)$ \\
16943689 &        & $09^{\rm h}40^{\rm m}51.03^{\rm s}$ & $+34^{\circ}52'05.2''$ & $ 0.3415961 (11)$ & $ 0.4598366 (25)$ \\
21957980 &        & $10^{\rm h}23^{\rm m}23.49^{\rm s}$ & $+26^{\circ}48'55.8''$ & $ 0.3600373 (15)$ & $ 0.4838723 (32)$ \\
22316675 &        & $10^{\rm h}30^{\rm m}08.33^{\rm s}$ & $+03^{\circ}36'08.2''$ & $ 0.4344241 (10)$ & $ 0.5812557 (43)$ \\
22657637 &        & $10^{\rm h}33^{\rm m}29.19^{\rm s}$ & $+03^{\circ}26'27.0''$ & $ 0.4150979 (13)$ & $ 0.5573046 (48)$ \\
22542640 &        & $10^{\rm h}39^{\rm m}41.62^{\rm s}$ & $+11^{\circ}18'18.2''$ & $ 0.4259760 (09)$ & $ 0.5716267 (37)$ \\
23148883 &        & $10^{\rm h}40^{\rm m}40.76^{\rm s}$ & $+09^{\circ}02'09.0''$ & $ 0.3901304 (11)$ & $ 0.4909742 (48)$ \\
23135759 &        & $10^{\rm h}41^{\rm m}58.19^{\rm s}$ & $+07^{\circ}12'42.1''$ & $ 0.4027333 (07)$ & $ 0.5398760 (25)$ \\
22922490 &        & $10^{\rm h}46^{\rm m}24.40^{\rm s}$ & $+20^{\circ}08'52.9''$ & $ 0.3389022 (10)$ & $ 0.4564827 (20)$ \\
23184879 &        & $10^{\rm h}50^{\rm m}38.73^{\rm s}$ & $+12^{\circ}27'27.6''$ & $ 0.3763188 (12)$ & $ 0.5054710 (21)$ \\
23378473 & AK UMa & $10^{\rm h}53^{\rm m}13.17^{\rm s}$ & $+41^{\circ}19'01.5''$ & $ 0.3658605 (13)$ & $ 0.4915826 (28)$ \\
23713423 &        & $10^{\rm h}59^{\rm m}57.15^{\rm s}$ & $+37^{\circ}50'20.2''$ & $ 0.4745222 (29)$ & $ 0.3526346 (12)$ \\
23500879 &        & $11^{\rm h}00^{\rm m}58.03^{\rm s}$ & $+01^{\circ}52'53.4''$ & $ 0.3548337 (07)$ & $ 0.4766697 (22)$ \\
23675270 &        & $11^{\rm h}01^{\rm m}03.50^{\rm s}$ & $+04^{\circ}28'21.6''$ & $ 0.3573184 (07)$ & $ 0.4799444 (21)$ \\
23767174 &        & $11^{\rm h}01^{\rm m}53.62^{\rm s}$ & $+06^{\circ}24'20.3''$ & $ 0.3688578 (05)$ & $ 0.4947679 (34)$ \\
23624228 &        & $11^{\rm h}02^{\rm m}37.79^{\rm s}$ & $+23^{\circ}14'15.3''$ & $ 0.3644500 (10)$ & $ 0.4895522 (23)$ \\
23968149 &        & $11^{\rm h}08^{\rm m}33.65^{\rm s}$ & $+21^{\circ}52'38.2''$ & $ 0.3784654 (11)$ & $ 0.5084678 (32)$ \\
24050838 &        & $11^{\rm h}10^{\rm m}21.80^{\rm s}$ & $+35^{\circ}46'50.4''$ & $ 0.3504849 (15)$ & $ 0.4710607 (55)$ \\
1080552  &        & $11^{\rm h}15^{\rm m}52.46^{\rm s}$ & $+10^{\circ}51'26.8''$ & $ 0.3514491 (08)$ & $ 0.4725646 (24)$ \\
1113422  &        & $11^{\rm h}21^{\rm m}26.07^{\rm s}$ & $+12^{\circ}34'47.8''$ & $ 0.3862376 (09)$ & $ 0.5178649 (29)$ \\
1435279  &        & $11^{\rm h}29^{\rm m}14.77^{\rm s}$ & $+01^{\circ}15'33.5''$ & $ 0.3818601 (09)$ & $ 0.5118339 (31)$ \\
1492054  &        & $11^{\rm h}31^{\rm m}17.44^{\rm s}$ & $+16^{\circ}39'56.4''$ & $ 0.3566548 (10)$ & $ 0.4791520 (23)$ \\
2122319  &        & $11^{\rm h}47^{\rm m}20.18^{\rm s}$ & $+01^{\circ}49'26.2''$ & $ 0.3594229 (08)$ & $ 0.4826001 (26)$ \\
2251043  &        & $11^{\rm h}51^{\rm m}41.66^{\rm s}$ & $-01^{\circ}08'49.2''$ & $ 0.3476471 (13)$ & $ 0.4679691 (19)$ \\
3026086  &        & $12^{\rm h}06^{\rm m}31.48^{\rm s}$ & $+07^{\circ}46'21.4''$ & $ 0.3981172 (08)$ & $ 0.5331859 (59)$ \\
3384231  &        & $12^{\rm h}21^{\rm m}38.87^{\rm s}$ & $+00^{\circ}48'59.2''$ & $ 0.3443139 (15)$ & $ 0.4633847 (33)$ \\
4282396  &        & $12^{\rm h}33^{\rm m}54.53^{\rm s}$ & $+08^{\circ}16'37.9''$ & $ 0.3922362 (06)$ & $ 0.5260051 (25)$ \\
7527829  &        & $12^{\rm h}43^{\rm m}22.85^{\rm s}$ & $+34^{\circ}57'16.7''$ & $ 0.3622289 (10)$ & $ 0.4866526 (35)$ \\
7827663  &        & $12^{\rm h}49^{\rm m}28.96^{\rm s}$ & $+03^{\circ}22'41.5''$ & $ 0.3908330 (08)$ & $ 0.5244291 (29)$ \\
7792311  & AZ Com & $12^{\rm h}53^{\rm m}50.13^{\rm s}$ & $+22^{\circ}18'39.6''$ & $ 0.3998350 (11)$ & $ 0.5361411 (44)$ \\
8119639  & ES Com & $12^{\rm h}59^{\rm m}07.46^{\rm s}$ & $+26^{\circ}35'50.2''$ & $ 0.4175065 (17)$ & $ 0.5598161 (67)$ \\
8222011  &        & $13^{\rm h}02^{\rm m}04.51^{\rm s}$ & $+46^{\circ}35'34.1''$ & $ 0.3509191 (10)$ & $ 0.4717867 (32)$ \\
8378112  & RR Com & $13^{\rm h}10^{\rm m}35.95^{\rm s}$ & $+18^{\circ}01'10.5''$ & $ 0.3805465 (23)$ & $ 0.5106514 (51)$ \\
8735344  & GG Com & $13^{\rm h}19^{\rm m}54.05^{\rm s}$ & $+29^{\circ}42'22.5''$ & $ 0.4033257 (12)$ & $ 0.5407703 (39)$ \\
9142645  &        & $13^{\rm h}26^{\rm m}27.27^{\rm s}$ & $+05^{\circ}17'45.8''$ & $ 0.3557081 (12)$ & $ 0.4780519 (41)$ \\
9005186  &        & $13^{\rm h}26^{\rm m}36.68^{\rm s}$ & $+21^{\circ}33'44.9''$ & $ 0.4042162 (13)$ & $ 0.5417068 (62)$ \\
9236215  &        & $13^{\rm h}31^{\rm m}22.43^{\rm s}$ & $-00^{\circ}22'40.3''$ & $ 0.3525714 (07)$ & $ 0.4739840 (21)$ \\
9984569  &        & $13^{\rm h}51^{\rm m}02.14^{\rm s}$ & $-01^{\circ}47'40.3''$ & $ 0.3487413 (11)$ & $ 0.4687500 (34)$ \\
13690975 &        & $15^{\rm h}04^{\rm m}12.33^{\rm s}$ & $+08^{\circ}18'41.6''$ & $ 0.3532253 (09)$ & $ 0.4750572 (22)$ \\
13459905 &        & $15^{\rm h}05^{\rm m}48.01^{\rm s}$ & $+22^{\circ}48'54.4''$ & $ 0.3785326 (13)$ & $ 0.5076959 (42)$ \\
13551583 &        & $15^{\rm h}07^{\rm m}26.75^{\rm s}$ & $+32^{\circ}43'46.7''$ & $ 0.4040864 (16)$ & $ 0.5418266 (49)$ \\
13905365 &        & $15^{\rm h}16^{\rm m}09.22^{\rm s}$ & $+32^{\circ}00'07.3''$ & $ 0.3499099 (14)$ & $ 0.4706344 (40)$ \\
15456241 &        & $15^{\rm h}47^{\rm m}40.81^{\rm s}$ & $+38^{\circ}29'16.5''$ & $ 0.3590427 (11)$ & $ 0.4821719 (20)$ \\
16200802 &        & $16^{\rm h}01^{\rm m}39.89^{\rm s}$ & $+03^{\circ}25'58.9''$ & $ 0.3563820 (09)$ & $ 0.4786143 (28)$ \\
17942645 &        & $16^{\rm h}17^{\rm m}29.01^{\rm s}$ & $+04^{\circ}20'34.0''$ & $ 0.4237921 (15)$ & $ 0.5686818 (52)$ \\
18021143 &        & $16^{\rm h}18^{\rm m}23.18^{\rm s}$ & $+43^{\circ}39'38.6''$ & $ 0.4583683 (22)$ & $ 0.3403732 (14)$ \\
19024518 &        & $16^{\rm h}45^{\rm m}06.49^{\rm s}$ & $+22^{\circ}38'27.6''$ & $ 0.3558346 (08)$ & $ 0.4780277 (44)$ \\
19533186 &        & $16^{\rm h}48^{\rm m}58.71^{\rm s}$ & $+43^{\circ}26'09.9''$ & $ 0.3569764 (19)$ & $ 0.4799628 (43)$ \\
\enddata
\end{deluxetable}

\begin{deluxetable}{lrrrrrr}
\tabletypesize{\footnotesize}
\tablecaption{RRd pulsators confirmed using EROS-II LMC data -- example part\label{tab:eros_newRRd}}
\tablewidth{0pt}
\tablehead{
\colhead{EROS-II ID} & \colhead{R.A.} &
\colhead{Dec.} & \colhead{$P_{\rm 1O}~[{\rm d}]$} & \colhead{$\sigma_{\rm P,1O}~[{\rm d}]$} & \colhead{$P_{\rm F}~[{\rm d}]$} & \colhead{$\sigma_{\rm P,F}~[{\rm d}]$} 
}
\startdata
lm0660l9646  & 4:44:19.7 & -72:03:41.9 & 0.40070696 & 0.00000197 & 0.53665880 & 0.00000951 \\
lm0536n10190 & 4:44:39.2 & -71:42:58.4 & 0.35705577 & 0.00000183 & 0.47961601 & 0.00000685 \\
lm0664l10948 & 4:45:26.1 & -72:46:32.4 & 0.36628044 & 0.00000127 & 0.49209884 & 0.00000317 \\
lm0660n13045 & 4:46:20.6 & -72:05:07.6 & 0.40663011 & 0.00000147 & 0.54503290 & 0.00000554 \\
lm0660n21286 & 4:46:39.2 & -72:08:42.5 & 0.36565261 & 0.00000137 & 0.49117710 & 0.00000358 \\
\multicolumn{7}{l}{\ldots} \\
\enddata
\tablecomments{Table \ref{tab:eros_newRRd} is published in its entirety ($350$ records) in the
electronic edition of the {\it PASP}.  A portion is
shown here for guidance regarding its form and content.}
\end{deluxetable}

\begin{deluxetable}{lrrrrrr}
\tabletypesize{\footnotesize}
\tablecaption{Double and triple mode pulsating stars other than RRd confirmed in EROS-II LMC data \label{tab:eros_other}}
\tablewidth{0pt}
\tablehead{
\colhead{EROS-II ID} & \colhead{R.A.} &
\colhead{Dec.} & \colhead{$P_{1}~[{\rm d}]$} & \colhead{$P_{2}~[{\rm d}]$} & \colhead{$P_{1}/P_{2}$} &
\colhead{$P_{3}~[{\rm d}]$}
}
\startdata
lm0192l19199\tablenotemark{a} & 5:15:14.7 & -68:16:19.7 & $0.36224380 (376)$ & $0.49720612 (726)$ & $0.72856$ & \\
lm0447m22639 & 5:16:25.2 & -66:03:26.1 & $0.36319182  (88)$  & $0.27398322 (94)$  & $0.75438$ & $0.15617123 (55)$ \\
lm0324m3996\tablenotemark{b}  & 5:18:51.3 & -66:58:32.1 & $0.37804446 (66)$  & $0.30405400 (73)$  & $0.80428$ &  \\
lm0707k13116\tablenotemark{c} & 5:25:47.9 & -73:02:15.0 & $0.37241075 (55)$  & $0.29968396 (151)$ & $0.80471$ & $0.49174877 (481)$ \\
lm0582n31174\tablenotemark{d} & 5:26:52.2 & -71:08:28.1 & $0.38536004 (140)$ & $0.51017721 (437)$ & $0.75535$ &  \\
lm0591l8196\tablenotemark{e}  & 5:38:29.1 & -70:39:21.0 & $0.39033675 (103)$ & $0.31337215 (271)$ & $0.80283$ &  \\
lm0604l26198\tablenotemark{f} & 5:43:43.4 & -71:29:03.5 & $0.37583649 (175)$ & $0.51482140 (356)$ & $0.73003$ &  \\
lm0723l27967 & 5:45:17.0 & -72:31:58.3 & $0.38084033 (171)$ & $0.50829032 (250)$ & $0.74926$ & $0.51298738 (627)$ \\
\enddata
\tablenotetext{a}{OGLE-III data for this star \citep[OGLE-LMC-RRLYR-08917][]{soszynski09rrlyrlmc} confirm existance of the secondary period.}
\tablenotetext{b}{\citet{marquette09} presented this stars as one of 1O/2O classical Cepheids (ID J051851-665831).}
\tablenotetext{c}{This object is triple-mode classical Cepheid. See discussion in Section~3.}
\tablenotetext{d}{Secondary period of this star (OGLE-LMC-RRLYR-15686) was already noted by \citet{soszynski09rrlyrlmc}.}
\tablenotetext{e}{The stars was identified in the OGLE-II catalog of the LMC RR Lyr stars \citep{soszynski03} as OGLE\_053829.10-703920.3.}
\tablenotetext{f}{OGLE-III data for this star \citep[OGLE-LMC-RRLYR-22167][]{soszynski09rrlyrlmc} confirm existance of the secondary period.}
\end{deluxetable}

\begin{deluxetable}{lrrrrlrl}
\tabletypesize{\scriptsize}
\rotate
\tablecaption{Corrections to LINEAR catalog of variable stars\label{tab:other}}
\tablewidth{0pt}
\tablehead{
\colhead{ID} & \colhead{R.A.} & \colhead{Dec.} & \colhead{$m$} & 
\colhead{$P_{\rm LINEAR}$} & \colhead{type} & \colhead{$P_{\rm new}$} & \colhead{type/comments} \\
\colhead{} & \colhead{} & \colhead{} & \colhead{$[{\rm mag}]$} & 
\colhead{$[{\rm d}]$} & \colhead{} & \colhead{$[{\rm d}]$} & \colhead{}
}
\startdata
\multicolumn{8}{l}{\ldots} \\
9183325 & 13:24:13.55 & +60:55:07.0 & 16.55 & -9.900000 &              other & 1.22085215 &  \\                                                      
9328902 & 13:35:49.76 & +26:55:16.7 & 16.24 & 0.051747  & SX\_Phe/delta\_Sct & 0.05174768 & SX\_Phe/delta\_Sct F/1O, secondary period: 0.04046822 d \\ 
9345642 & 13:29:00.73 & +29:48:38.6 & 16.81 & 0.063909  & SX\_Phe/delta\_Sct & 0.06006094 & SX\_Phe/delta\_Sct                                      \\
9473742 & 13:30:50.76 & +54:07:45.6 & 14.49 & 0.063307  & SX\_Phe/delta\_Sct & 0.06330779 & SX\_Phe/delta\_Sct                                      \\
9634973 & 13:42:20.40 & +27:53:51.7 & 15.81 & -9.900000 &              other & 0.50936354 & RRab                      \\
\multicolumn{8}{l}{\ldots} \\
\enddata
\tablecomments{Table \ref{tab:other} is published in its entirety ($142$ records) in the
electronic edition of the {\it PASP}.  A portion is
shown here for guidance regarding its form and content.}
\end{deluxetable}

\end{document}